\documentclass[11pt]{article}
\usepackage{geometry}
\usepackage{times}
\usepackage[final]{graphicx}
\usepackage{mathrsfs}
\usepackage{amssymb}
\usepackage{amsmath}

\title{Capacity of the Degraded Half-Duplex Relay Channel\thanks{This
    research is supported in part by the National Science Foundation
    under Grant CNS-0626863.}}

\author{Saravanan Vijayakumaran$^\dag$, Tan F. Wong$^\dag$, and Tat M.
  Lok$^\ddag$}

\date{$\dag$Department of Electrical and Computer Engineering \\
  University of Florida \\
  Gainesville, FL 32611-6130, U.S.A. \\
  Email: \texttt{sarva.v@gmail.com,twong@ufl.edu} \\~\\
  $\ddag$Department of Information Engineering \\
  The Chinese University of Hong Kong \\
  Shatin, Hong Kong \\
  Email: \texttt{tmlok@ie.cuhk.edu.hk}}

\begin{document}
\maketitle

\begin{abstract}
  A discrete memoryless half-duplex relay channel is constructed from
  a broadcast channel from the source to the relay and destination and
  a multiple access channel from the source and relay to the
  destination. When the relay listens, the channel operates in the
  broadcast mode. The channel switches to the multiple access mode
  when the relay transmits. If the broadcast component channel is
  physically degraded, the half-duplex relay channel will also be
  referred to as physically degraded. The capacity of this degraded
  half-duplex relay channel is examined. It is shown that the block
  Markov coding suggested in the seminal paper by Cover and El Gamal
  can be modified to achieve capacity for the degraded half-duplex
  relay channel. In the code construction, the listen-transmit
  schedule of the relay is made to depend on the message to be sent
  and hence the schedule carries information itself. If the schedule
  is restricted to be deterministic, it is shown that the capacity can
  be achieved by a simple management of information flows across the
  broadcast and multiple access component channels.
\end{abstract}

\section{Introduction}
The half-duplex relay channel differs from the full-duplex relay
channel \cite{Cover79} in the inability of the relay to simultaneously
transmit and receive signals. In many practical systems, the
half-duplex assumption provides a more realistic model of the relay
channel. A number of information theoretic studies
\cite{Gastpar02}--\cite{Kho03} of the half-duplex relay channel have
been carried out. The focus of these studies is usually on the special
case in which the links in the relay channel are Gaussian. In
particular, the half-duplex relay channel is modeled in
\cite{Kramer04} as a special case of a full-duplex relay channel with
an additional input symbol at the relay to describe whether the relay
is listening or transmitting. Then the bounds and achievability
results for the full-duplex relay channel in \cite{Cover79} are
directly applied to the half-duplex specialization. One interesting
observation made in \cite{Kramer04} is that the listen-transmit
schedule of the relay can actually carry additional information when
it is made to depend on the message to be sent. If the relay
listen-transmit schedule is restricted to be deterministic, i.e., the
same for any message, separate consideration, as suggested in
\cite{Host02,Host05}, of the time during which the relay listens or
transmits leads to bounds and achievability results for the
half-duplex relay channel that are similar to the corresponding
counterparts of the full-duplex relay channel.

In this paper, we consider a discrete memoryless half-duplex relay
channel.  Based on physical reasoning, the half-duplex relay channel
has two modes of operation --- the broadcast (BC) mode and the
multiple access (MA) mode. When the relay listens, the channel is in
the BC mode, which is specified by a BC channel from the source to the
relay and destination. When the relay transmits, the channel is in the
MA mode, which is specified by a MA channel from the source and relay
to the destination. Here we are mainly interested in the special case
in which the BC channel is physically degraded. By combining the
aforementioned ideas in \cite{Kramer04} and \cite{Host02,Host05}, we
show that the capacity of this degraded half-duplex relay channel can
be achieved via a decode-forward approach employing a modified version
of the block Markov coding technique suggested in \cite{Cover79} to
accommodate the inclusion of random relay listen-transmit schedule and
separate treatment of the two modes of the relay channel. If the relay
schedule is restricted to be deterministic, we argue that a simple
management of the flows of information across the BC and MA component
channels is sufficient to achieve capacity.

\section{A Model for the Half-Duplex Relay Channel}
We consider the half-duplex relay channel as a special case of the
classical discrete memoryless relay channel studied in \cite{Cover79}
with the addition of an extra input at the relay to describe its
state. This relay channel model as well as its specialization to the
half-duplex case will be described below. We note that a similar
formulation for the Gaussian relay channel is suggested in
\cite{Kramer04}.

\subsection{Channel model with relay state}
\begin{figure}
\begin{center}
\includegraphics[width=0.8\textwidth]{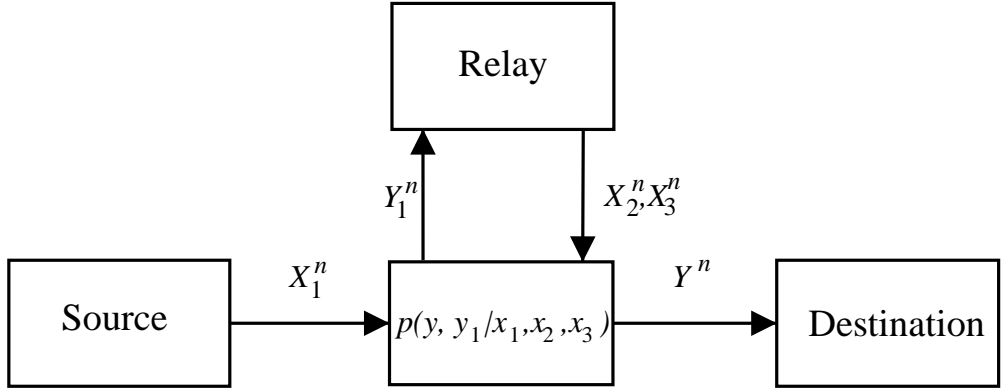}
\end{center}
\caption{The relay channel model with relay state.}
\label{fig:fdrelay}
\end{figure}

Consider the relay channel shown in Fig.~\ref{fig:fdrelay}.
It consists of five finite sets
$\mathcal{X}_1,\mathcal{X}_2,\mathcal{X}_3,\mathcal{Y},\mathcal{Y}_1$
and a collection of probability mass functions (pmfs)
$p(y,y_1|x_1,x_2,x_3)$, one for every $(x_1,x_2,x_3) \in \mathcal{X}_1
\times \mathcal{X}_2 \times \mathcal{X}_3$, where $(y,y_1) \in
\mathcal{Y} \times \mathcal{Y}_1$.  Here $\mathcal{X}_1$ denotes the
input alphabet at the source, $\mathcal{X}_2$ and $\mathcal{X}_3$
respectively denote the input alphabet and state at the relay,
$\mathcal{Y}$ denotes the output alphabet at destination and
$\mathcal{Y}_1$ denotes the output alphabet at the relay.
 
A $(2^{nR},n)$ code for the above relay channel consists of a set of
integers $\mathcal{W} = \{1,2,\ldots,2^{nR}\}$, an encoding function
$f:\mathcal{W} \rightarrow \mathcal{X}_1^n$, a collection of $n$
relaying functions $g_i:\mathcal{Y}_1^{i-1} \rightarrow \mathcal{X}_2
\times \mathcal{X}_3, \ \ \ \ 1 \leq i \leq n$, and a decoding
function $h: \mathcal{Y}^n\rightarrow \mathcal{W}$.  The relaying
functions are defined so that $i$th input symbol at the relay depends
only on the previously received output symbols at the relay.
%This description of the relay channel and its corresponding code
%differs from the traditional definitions in the availability of an
%extra input variable $X_3$ at the relay.  This variable enables a
%simple description of the half-duplex relay channel as a special case
%of the full-duplex relay channel.
If we assume that the distribution of messages over $\mathcal{W}$ is
uniform, the average probability of error for the $(2^{nR},n)$ code is
\[
P_e^{(n)} = \frac{1}{2^{nR}}\sum_{w\in\mathcal{W}} \Pr\{h(Y^n) \neq
w| w \mbox{ was sent} \}.
\]
A rate $R$ is achievable if there exist a sequence of $(2^{nR},n)$
codes with $P_e^{(n)} \rightarrow 0$.

\subsection{Specialization to the half-duplex relay channel}
\label{sec:model}
\begin{figure}
\begin{center}
\includegraphics[width=0.9\textwidth]{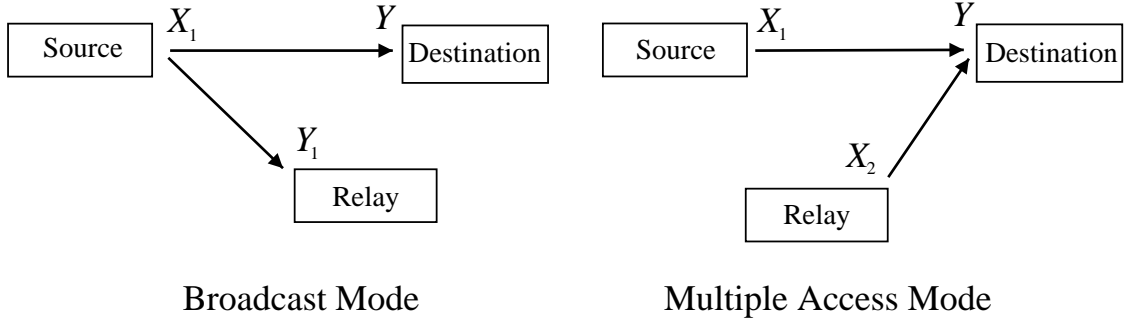}
\end{center}
\caption{The two modes of operation of a half-duplex relay channel}
\label{fig:modes}
\end{figure}

The relay in the half-duplex relay channel can either listen or
transmit, but not both at the same time. As a result, the half-duplex
relay channel has two modes of operation, the broadcast (BC) mode and
the multiple access (MA) mode, as illustrated in Fig.~\ref{fig:modes}.
In the BC mode, the channel is specified by
$(\mathcal{X}_1,p_{l}(y,y_1|x_1),\mathcal{Y},\mathcal{Y}_1)$, which
describes a BC channel from the source to the relay and destination.
Here $\mathcal{X}_1$ denotes the input alphabet at the source,
$\mathcal{Y}$ denotes the output alphabet at destination and
$\mathcal{Y}_1$ denotes the output alphabet at the relay.  The
collection of pmfs $p_{l}(y,y_1|x_1)$, one each for every $x_1 \in
\mathcal{X}_1$, where $(y,y_1) \in \mathcal{Y} \times \mathcal{Y}_1$,
specify the operation of this BC channel.  In the MA mode, the channel
is specified by
$(\mathcal{X}_1,\mathcal{X}_2,p_{t}(y|x_1,x_2),\mathcal{Y})$, which
describes a MA channel from the source and relay to the destination.
Here $\mathcal{X}_1$ denotes the input alphabet at the source,
$\mathcal{X}_2$ denotes the input alphabet at relay and $\mathcal{Y}$
denotes the output alphabet at the destination. The collection of pmfs
$p_{t}(y|x_1,x_2)$, one each for every $(x_1,x_2) \in \mathcal{X}_1
\times \mathcal{X}_2$, where $y \in \mathcal{Y} $, specify the
operation of this MA channel.

To account for the above two different modes of operation, we embed
the half-duplex relay channel in the channel model of the previous
subsection.  First, we set $\mathcal{X}_3 = \{l,t\}$, where $l$ will
denote the event when the relay is listening and $t$ will denote the
event when the relay is transmitting. The relay by listening or
transmitting causes the half-duplex relay channel to operate in the BC
mode and MA mode, respectively.  Without loss of generality, we can
assume that the output alphabet at the relay $\mathcal{Y}_1$ contains
an erasure symbol $e$ such that $p_l(y,e|x_1) = 0$ for all $(x_1,y)
\in \mathcal{X}_1 \times \mathcal{Y}$.  Then the channel pmf
$p(\cdot)$ of the previous subsection is defined in terms of
$p_{l}(\cdot)$ and $p_{t}(\cdot)$ in the following manner:
\begin{equation}
p(y,y_1|x_1,x_2,x_3)  =
  \left\{
    \begin{array}{ll}
      p_l(y,y_1|x_1) & \mbox{if $x_3 = l$} \\
      p_t(y|x_1,x_2) \delta_e(y_1) & \mbox{if $x_3 = t$}
    \end{array}
  \right.
\label{eqn:hdrpmf}
\end{equation}
for all $(x_1,x_2,y,y_1) \in \mathcal{X}_1 \times \mathcal{X}_2 \times
\mathcal{Y} \times \mathcal{Y}_1$, where $\delta_e(y_1) = 1$ if
$y_1=e$ and $\delta_e(y_1) = 0$ otherwise. This definition imposes the
physical restrictions that the outputs at the destination and relay in
the BC mode (when relay listens) do not depend on the relay input
$X_2$, and that the output at the relay is always erased in the MA
mode (when relay transmits).  In addition, we need the following
restriction on the source distribution $p(x_1,x_2,x_3)$ to accurately
reflect the operation of a half-duplex channel:
\begin{equation}
p(X_2 = x_2|X_3 = l)  = \delta_q (x_2)
\label{eqn:src_res}
\end{equation}
for some symbol $q \in \mathcal{X}_2$. The symbol $q$ can be thought
of as a ``quiet'' symbol chosen by the relay when it is listening. We
will implicitly assume that this restriction is satisfied whenever the
source distribution $p(x_1,x_2,x_3)$ is referred hereafter.

For the rest of the paper, we will assume that the BC channel
component is physically degraded, i.e., $p_l(y,y_1|x_1) = p_l(y|y_1)
p_l(y_1|x_1)$. If this condition is satisfied, we will call the
half-duplex relay channel physically degraded. We note that
however the channel $p(y,y_1|x_1,x_2,x_3)$ is not physically degraded
in the sense of \cite{Cover79}, i.e., $p(y,y_1|x_1,x_2,x_3) \neq
p(y|y_1,x_2,x_3) p(y_1|x_1,x_2,x_3)$.

\section{Outer Bound on the Capacity}
\label{s:ub}
The max-flow min-cut bound \cite[Thm. 14.10.1]{Cover91} on the relay
channel $p(y,y_1|x_1,x_2,x_3)$ gives the following outer bound:
\begin{eqnarray}
C \leq \max_{p(x_1,x_2,x_3)}\min
 \left\{I(X_1;Y,Y_1|X_2,X_3),I(X_1,X_2,X_3;Y)\right\}.
\label{eqn:maxmin}
\end{eqnarray}
Specializing this bound to the half-duplex relay channel, the first
term on the right hand side of (\ref{eqn:maxmin}) can be expanded as
follows,
\begin{eqnarray}
I(X_1;Y,Y_1|X_2,X_3)
& = & I(X_1;Y,Y_1|X_2,X_3 = l)p(X_3 =l) 
      + I(X_1;Y,Y_1|X_2,X_3 =t)p(X_3 =t) \nonumber \\
& = & I(X_1;Y,Y_1|X_3 =l)p(X_3 =l) +
      I(X_1;Y|X_2,X_3 =t)p(X_3 =t)
%& = & I(X_1;Y_1|X_3 =l)p(X_3 =l) + I(X_1;Y|X_2,X_3 =t)p(X_3 =t)
\label{eqn:halfbnd1}
\end{eqnarray}
where the second equality follows from the specialization in
(\ref{eqn:hdrpmf}) of the BC mode and the erasure of $Y_1$ in the MA
mode.  The second term on the right hand side of (\ref{eqn:maxmin})
can be written as
\begin{eqnarray}
I(X_1,X_2,X_3;Y) %& = & I(X_2,X_3;Y) + I(X_1;Y|X_2,X_3) \nonumber\\
  %& = & I(X_2,X_3;Y) + I(X_1;Y|X_2,X_3 = l)\Pr(X_3 =l) + I(X_1;Y|X_2,X_3 = t) \Pr(X_3 =t) \nonumber \\
  %& = & I(X_2,X_3;Y) + I(X_1;Y|X_3 = l)p(X_3 =l) + I(X_1;Y|X_2,X_3 = t)p(X_3 =t) \nonumber \\
  & = & I(X_3;Y) + I(X_1,X_2;Y|X_3=t)p(X_3=t) + I(X_1,X_2;Y|X_3 = l)p(X_3 =l) \nonumber \\
  & = & I(X_3;Y) + I(X_1,X_2;Y|X_3=t)p(X_3=t) + I(X_1;Y|X_3 = l)p(X_3 =l)
\label{eqn:halfbnd2}
\end{eqnarray}
where the second equality is due to the fact that $I(X_1;Y|X_2,X_3=l) =
I(X_1;Y|X_3=l)$ and $I(X_2;Y|X_3=l) = 0$.  Substituting
(\ref{eqn:halfbnd1}) and (\ref{eqn:halfbnd2}) in (\ref{eqn:maxmin}),
we get
\begin{eqnarray}
C & \leq &
 \max_{p(x_1,x_2,x_3)} \Big\{ I(X_1;Y|X_2,X_3 = t) p(X_3 =t)
    + \min\big\{I(X_1;Y,Y_1|X_3 =l)p(X_3 =l), \nonumber \\
    & & ~~~~~~~~~~~~~I(X_3;Y)+I(X_2;Y|X_3=t)p(X_3 =t) +
    I(X_1;Y|X_3 = l)p(X_3 =l)\big\} \Big\} 
\label{eqn:halfbndout} \\
&=& 
 \max_{p(x_1,x_2,x_3)} \Big\{ I(X_1;Y|X_2,X_3 = t) p(X_3 =t)
    + \min\big\{I(X_1;Y_1|X_3 =l)p(X_3 =l), \nonumber \\
    & & ~~~~~~~~~~~~~I(X_3;Y)+I(X_2;Y|X_3=t)p(X_3 =t) +
    I(X_1;Y|X_3 = l)p(X_3 =l)\big\} \Big\}
\label{eqn:halfbndfin}
\end{eqnarray}
where the equality in the last line above is due to the degradedness
of the half-duplex relay channel.

\section{Achievability}
\label{s:achiev}
The upper bound on capacity in (\ref{eqn:halfbndfin}) suggests a
particular structure for the optimal code.
%Typical achievability results which use a random coding argument
%involve an upper bound on the probability of error which often has the
%form $2^{n(R-I)}$, where $R$ is the rate of the code and $I$ is a
%mutual information between certain input and output variables. This
%bound decreases to zero as $n\rightarrow \infty$, if $R<I$. In most
%cases, the $2^{nR}$ term corresponds approximately to the number of
%codewords in the code and the $2^{-nI}$ term corresponds approximately
%to the probability that a pair of independently generated random
%variable sequences of length $n$ are jointly typical.
%%
We note that the upper bound involves products of conditional mutual
informations, where the conditioning is on particular values of $X_3$,
and the probabilities of $X_3$ taking these values.  Such products can
arise in the bounds on the probability of error, if we restrict our
attention to those coordinates of the $n$-sequences of random
variables where $X_3$ takes a particular value. Suppose we generate a
random code and restrict the code to the coordinates where $X_3 = t$.
Then the probability that a pair of independently generated random
variable sequences are jointly typical will be approximately equal to
$2^{-a_n I}$, where $a_n$ is the number of coordinates where $X_3 =t$
and $I$ is a conditional mutual information conditioned on $X_3 =t$.
Since $\frac{a_n}{n} \rightarrow p(X_3 =t)$ as $n \rightarrow \infty$,
for large $n$ this probability can be approximated as $2^{-n(I p(X_3
  =t))}$.  This argument suggests that a random code which employs
joint typicality decoding can hope to achieve the outer bound if it
treats the blocks of coordinates corresponding to different values of
$X_3$ differently. With this observation, the block Markov coding in
\cite{Cover79} can be employed to construct a code with rate achieving
the outer bound in the previous section.

\subsection{Random Code Construction}
Fix a source distribution $p(x_1,x_2,x_3)$ under the restriction of
(\ref{eqn:src_res}) with $\mathcal{X}_3 = \{l,t\}$. Further, assume
that $0<p(X_3 =l)<1 $. If $p(X_3 =l)$ is $0$ or $1$, the bound in
(\ref{eqn:halfbndfin}) takes a very simple form which can be achieved
by the usual random coding argument.

Generate $2^{nR_4}$ independent identically distributed (iid)
$n$-sequences $\mathbf{x}_3(s) \in \mathcal{X}_3^n$, \\
$s \in \{1,2,\ldots,2^{nR_4}\}$, each drawn according to
$p(\mathbf{x}_3) = \prod_{i=1}^{n} p(x_{3i})$.

For each $\mathbf{x}_3(s)$, generate $2^{nR_3}$ conditionally
independent $n$-sequences $\mathbf{x}_2(w|s) \in \mathcal{X}_3^n$, $w
\in \{1,2,\ldots,2^{nR_3}\}$, each one drawn according to
$p(\mathbf{x}_2|\mathbf{x}_3(s) = \prod_{i=1}^{n}
p(x_{2i}|x_{3i}(s))$.

For each $\mathbf{x}_3(s)$, define the index sets $A_n(s) = \{1\leq
i\leq n:\ x_{3i}(s) =t\}$ and $B_n(s) = \{1\leq i\leq n :\ x_{3i}
=l\}$.  For each pair $(\mathbf{x}_2(w),\mathbf{x}_3(s))$, we wish to
generate $2^{n(R_1 + R_2)}$ $n$-sequences $\mathbf{x}_1(u,v|w,s) \in
\mathcal{X}_1^n$, $u\in\{1,2,\ldots,2^{nR_1}\}$, $v \in
\{1,2,\ldots,2^{nR_2}\}$ in the following manner. For those indices in
$A_n(s)$, generate $2^{nR_1}$ conditionally independent
$|A_n(s)|$-sequences in $\mathcal{X}_1^{|A_n(s)|}$ each drawn
according to
\[
p(\{x_{1i_k}:i_k \in A_n(s)\}|\mathbf{x}_2(w),\mathbf{x}_3(s)) =
\prod_{k=1}^{|A_n(s)|} p(x_{1i_k}|x_{2i_k}(w),x_{3i_k}(s)=t).
\]
Index each such sequence by $u \in \{1,2,\ldots,2^{nR_1}\}$.
Similarly, for the indices in $B_n(s)$, generate $2^{nR_2}$
conditionally independent $|B_n(s)|$-sequences in
$\mathcal{X}_1^{|B_n(s)|}$ each drawn according to
\[
p(\{x_{1i_k}:i_k \in B_n(s)\}|\mathbf{x}_2(w),\mathbf{x}_3(s)) =
\prod_{k=1}^{|B_n(s)|} p(x_{1i_k}|x_{2i_k}(w),x_{3i_k}(s)=l).
\]
Index each such sequence by $v \in \{1,2,\ldots,2^{nR_2}\}$.  Thus for
each pair $(w,s)$, a pair of indices $(u,v) \in
\{1,2,\ldots,2^{nR_1}\} \times \{1,2,\ldots,2^{nR_2}\}$ specifies an
$n$-sequence $\mathbf{x}_1(u,v|w,s) \in \mathcal{X}_1^n$.

Let $R_0 = R_3 +R_4$. Generate a random partition $\mathcal{S} =
\{S_1,S_2,\ldots,S_{2^{nR_0}}\}$ of $\{1,2,\ldots,2^{nR_2}\}$ by
assigning each $v \in \{1,2,\ldots,2^{nR_2}\}$ independently to cell
$S_i$, according to a uniform distribution over the indices
$i=1,2,\ldots,2^{nR_0}$.  Note that there is a one-to-one
correspondence between the set of partition indices and the set
$\{1,2,\ldots,2^{nR_3}\} \times \{1,2,\ldots,2^{nR_4}\}$.  Thus, each
paritition element $S_i$ can alternatively be written as $S_{w,s}$
where $(w,s)$ is a unique pair in $\{1,2,\ldots,2^{nR_3}\} \times
\{1,2,\ldots,2^{nR_4}\}$.

\subsection{Encoding Scheme}
Let $R=R_1+R_2$.  We will use $B$ blocks, each having $n$ symbols, to
send a sequence of $B-1$ messages $m_i \in \{1,2,\ldots,2^{nR}\}$,
$i=1,2,\ldots,B-1$.  As $B\rightarrow \infty$, the rate
$\frac{R(B-1)}{B}$ will get arbitrarily close to $R$.

We assume that the messages $m_i$ are generated according to a uniform
distribution on the set $\{1,2,\ldots,2^{nR}\}$. These messages can be
easily mapped on to the set $\{1,2,\ldots,2^{nR_1}\} \times
\{1,2,\ldots,2^{nR_2}\}$, to generate random variables $u_i$ and $v_i$
which are uniformly distributed on the sets $\{1,2,\ldots,2^{nR_1}\} $
and $ \{1,2,\ldots,2^{nR_2}\}$, respectively.

The transmission scheme is as follows. For each message $m_i$ to be
sent in block $i$, calculate the variables $u_i$ and $v_i$. Let
$v_{i-1} \in S_{w_i,s_i}$. Then the source sends the codeword
$\mathbf{x}_1(u_i,v_i|w_i,s_i)$ in block $i$. The relay estimates
$v_{i-1}$ as $\hat{v}_{i-1}$ from the transmission in the previous
block. The estimation procedure will be explained in the next section.
Let $\hat{v}_{i-1} \in S_{\hat{w}_i,\hat{s}_i}$. Then the relay sends
the codeword $(\mathbf{x}_2(\hat{w}_i),\mathbf{x}_3(\hat{s}_i))$ in
block $i$.  For the first block, the relay does not have a previous
message to estimate. In this case, any message can be used as long as
the source is aware of it.

\subsection{Decoding Scheme}
\label{sec:decsch}
We will assume that at the end of the transmission of block $i-1$ the
destination knows $(u_1,\ldots,u_{i-1})$, $(v_1,\ldots,v_{i-2})$ and
$((w_1,s_1),\ldots,(w_{i-1},s_{i-1}))$. We also assume that the relay
knows $(v_1,\ldots,v_{i-1})$ and consequently knows
%knows $((u_1,v_1),(u_2,v_2),\ldots,(u_{i-1},v_{i-1}))$ and 
$((w_1,s_1),(w_2,s_2),\ldots,(w_{i},s_{i}))$, since the latter is a
function of the former.

At the end of block $i$, the following decoding operations are carried
out:
\begin{enumerate}
  
\item Upon receiving the $i$th block of outputs $\mathbf{Y}_1(i)$, and
  knowing $(w_i,s_i)$, the relay isolates those coordinates $j$ of the
  output where $x_{3j}(s_i) =l$ as $\mathbf{Y}^l_1(i)$, which is a
  sequence of $|B_n(s_i)|$ symbols. The same operation on the inputs
  gives us the $|B_n(s_i)|$-sequences
  $\mathbf{x}_1^l(u_i,v_i|w_i,s_i)$, $\mathbf{x}_2^l(w_i)$ and
  $\mathbf{x}_3^l(s_i)$.  Due to the structure of the code,
  $\mathbf{x}_1^l(u_i,v_i|w_i,s_i)$ does not depend on $u_i$ and can
  be written as $\mathbf{x}_1^l(v_i|w_i,s_i)$.  The relay declares
  that $\hat{v}_i = v$ was sent by the source in the $i$th block if
  and only if there exists a unique $v \in \{1,2,\ldots,2^{nR_2}\}$
  such that $(\mathbf{x}_1^l(v|w_i,s_i),\mathbf{x}_2^l(w_i),
  \mathbf{x}_3^l(s_i),\mathbf{Y}_1^l(i))$ are jointly typical.  This
  constitutes the estimation procedure at the relay used in the
  encoding scheme.  Note that $\hat{v}_i = v_i$ with arbitrarily small
  probability of error if
\begin{equation}
R_2 < I(X_1;Y_1|X_3=l)p(X_3 =l)
\label{eqn:achR_21}
\end{equation}
and the blocklength $n$ is sufficiently large.

\item Upon receiving the $i$th block of outputs $\mathbf{Y}(i)$, the
  destination declares that $(\hat{w}_i,\hat{s}_i) = (w,s)$ was sent
  by the relay in the $i$th block if and only if there exists a unique
  pair $(w,s) \in \{1,2,\ldots,2^{nR_3}\} \times
  \{1,2,\ldots,2^{nR_4}\}$ such that
  $(\mathbf{x}_2(w),\mathbf{x}_3(s),\mathbf{Y}(i))$ are jointly
  typical. From the properties of jointly typical sequences,
  $(\hat{w}_i,\hat{s}_i) = (w_i,s_i)$ with arbitrarily small
  probability of error if
\begin{equation}
R_0 < I(X_3;Y) + I(X_2;Y|X_3=t)p(X_3=t)
\label{eqn:achR_0}
\end{equation}
and the blocklength $n$ is sufficiently large.
 
\item Assuming that $(w_i,s_i)$ is correctly decoded in the previous
  step, the destination isolates those coordinates $j$ of the $i$th
  block output $\mathbf{Y}(i)$ where $x_{2j}(s_i) = t$, as
  $\mathbf{Y}^t(i)$, which is a sequence of $|A_n(s_i)|$ symbols.  The
  same operation on the inputs gives us the $|A_n(s_i)|$-sequences
  $\mathbf{x}_1^t(u_i,v_i|w_i,s_i)$, $\mathbf{x}_2^t(w_i)$ and
  $\mathbf{x}_3^t(s_i)$. As before, the structure of the code permits
  us to write $\mathbf{x}_1^t(u_i,v_i|w_i,s_i)$ as
  $\mathbf{x}_1^t(u_i|w_i,s_i)$.  The destination declares that
  $\hat{u}_i = u$ was sent by the source in the $i$th block if and
  only if there exists a unique $u \in \{1,2,\ldots,2^{nR_1}\}$ such
  that $(\mathbf{x}_1^t(u|w_i,s_i),\mathbf{x}_2^t(w_i),
  \mathbf{x}_3^t(s_i),\mathbf{Y}^t(i))$ are jointly typical. Again
  $\hat{u}_i = u_i$ with arbitrarily small probability of error if
\begin{equation}
R_1 < I(X_1;Y|X_2,X_3=t)p(X_3 =t) 
\label{eqn:achR_1}
\end{equation}
and the blocklength $n$ is sufficiently large.

\item Although, the destination does not know $v_{i-1}$ at the end of
  block $i-1$, it calculates an ambiguity set
  $\mathcal{L}(\mathbf{Y}(i-1))$, which consists of the set of all $v$
  such that $(\mathbf{x}_1^l(v|w_{i-1},s_{i-1}),
  \mathbf{x}_2^l(w_{i-1}),$
  $\mathbf{x}_3^l(s_{i-1}),\mathbf{Y}^l(i-1))$ are jointly typical.
  Assuming that $(w_i,s_i)$ is correctly decoded, the destination
  declares $\hat{v}_{i-1} = v$ was sent by the source in block $i-1$
  if and only if there exists a unique $v \in S_{w_i,s_i} \cap
  \mathcal{L}(\mathbf{Y}(i-1))$.  Like before, $\hat{v}_{i-1} =
  v_{i-1}$ with arbitrarily small probability of error if
\begin{equation}
R_2 < I(X_1;Y|X_3 =l)p(X_3=l) + R_0
\label{eqn:achR_22}
\end{equation} 
and $n$ is sufficiently large.
\end{enumerate}
The validity of the vanishing error probability claims above can be
verified by a detailed calculation. The approach is essentially the
same as the one given in \cite{Cover79} with the exception that strong
typicality is needed to ensure the convergence of the ratios $a_n/n$
and $b_n/n$ to $p(X_3=t)$ and $p(X_3=l)$, respectively. The details
are omitted here due to page limitation.  As the above decoding
operations at the end of block $i$ are successful, the destination
knows $u_i$ and $v_{i-1}$ in addition to what it knew at the end of
block $i-1$. Similarly, the relay knows $v_i$ and hence
$(w_{i+1},s_{i+1})$ in addition to its previous knowledge. Thus
correct decoding at each step ensures that this recursive decoding
scheme can proceed unhindered.

Since $R=R_1+R_2$, from (\ref{eqn:achR_21})--(\ref{eqn:achR_22}), we
have that $R$ is an achievable rate for the half-duplex relay channel
if
\begin{eqnarray}
R & < &  
  \underbrace{I(X_1;Y|X_3 = l)}_{R_1^l} p(X_3 =l)
   + \underbrace{I(X_1;Y|X_2,X_3 = t)}_{R_1^t} p(X_3 =t)
  \nonumber \\
& & 
  + \min\big\{[
  \underbrace{I(X_1;Y_1|X_3 =l) - I(X_1;Y|X_3 = l)}_{R_2^l}]
  p(X_3 =l), \nonumber \\
& & ~~ 
  I(X_3;Y)+ \underbrace{I(X_2;Y|X_3=t)}_{R_2^t} p(X_3=t)
  \big\}.
\label{eqn:halfach}
\end{eqnarray}
This achievable rate coincides with the outer bound in
(\ref{eqn:halfbndfin}) and hence gives the capacity of the physically
degraded half-duplex relay channel. 

\section{Random vs. Deterministic Relay Listen-Transmit Schedule}
\label{sec:randvsfixed}
With the relay state $X_3$ included as part of the source, we have
implicitly assumed that the listen-transmit schedule of the relay
depends on the message to be sent. Thus the schedule itself carries
information.  This is evidenced by the existence of the term
$I(X_3;Y)$ in (\ref{eqn:halfach}).  The use of random relay
listen-transmit schedules to increase the transmission rate in the
half-duplex relay channel was first suggested in \cite{Kramer04}. In
\cite{Kramer04}, the block Markov coding of \cite{Cover79} is directly
applied to the vector-valued input at the relay consisting of $X_2$
and $X_3$.  Since there is no point of sending information from the
source to the relay when the relay is transmitting, the source should
use all its resources to send to the destination. The vector-valued
relay input approach of \cite{Kramer04} is not able to do so as it
does not separate the two modes of operation of the half-duplex relay
channel. As a result, it can not achieve capacity.

If we restrict the relay listen-transmit schedule to be deterministic,
i.e., the schedule does not depend on the message to be sent, an
argument similar to the one in the previous two sections can be used
to show that the capacity of the degraded half-duplex relay channel is
given by the bound in (\ref{eqn:halfbndfin}) with the term $I(X_3;Y)$
removed and $p(X_3=t)$ and $p(X_3=l)$ interpreted as the fractions of
time the relay transmits and listens, respectively. Compared to the
random schedule, the loss of fixing the relay listen-transmit schedule
is $I(X_3;Y)$, which is no larger than $1$ bit. Block Markov coding
was proposed in \cite{Host02} to achieve the capacity of the Gaussian
half-duplex relay channel with deterministic relay schedule.  It is
also worth mentioning that the capacity can also be achieved using a
coding design that treats the received symbols at the destination
during the two modes of operation of the half-duplex relay channel as
those from a pair of parallel Gaussian channels \cite{Host05}. The
advantage of the latter approach is that all decoding can be done
within a single block.

Let $\alpha = p(X_3=l)$ be the fraction of time that the relay
listens. Then achievable rate of the half-duplex relay channel with
deterministic relay listen-transmit schedule can be written as
\begin{equation}
R < \alpha R_1^l + (1-\alpha) R_1^t + \min\{ \alpha R_2^l, (1-\alpha)R_2^t\}
\label{eqn:fixed}
\end{equation}
where $R_1^1$, $R_1^t$, $R_2^l$, and $R_2^t$ are given in
(\ref{eqn:halfach}).  A closer inspection of their corresponding
mutual information terms reveals an interesting interpretation. As
mentioned before, the half-duplex relay channel is made up of the BC
and MA components. The rate pair $(R_1^l,R_2^l)$ can be viewed as the
rate pair achievable over the (degraded) BC channel (c.f.
\cite[Thm.~14.6.2]{Cover91}), where $R_1^l$ is the rate of the flow of
information from the source to the destination and $R_2^l$ is the rate
of another distinct flow of information to the relay. Similarly the
rate pair $(R_1^t,R_2^t)$ represents the pair achievable over the MA
channel (c.f. \cite[Thm. 14.3.1]{Cover91}), where $R_1^t$ and $R_2^t$
are the rates from the source and the relay to the destination,
respectively. Since the relay is neither a source nor sink of
information, the $\min$ operator in (\ref{eqn:fixed}) specifies the
max-flow through the relay. This interpretation suggests another
simple way to achieve capacity in the case of physically degraded
half-duplex relay channel with deterministic relay listen-transmit
schedule:
\begin{enumerate}
\item Divide available time into 2 slots of fractional durations
  $\alpha$ and $1-\alpha$, respectively. The relay listens in the
  first time slot and transmits in the second one.
\item During the first time slot, the source broadcasts two distinct
  flows of information with rates $\alpha R_1^l$ and $\min\{ \alpha
  R_2^l, (1-\alpha)R_2^t\}$ to the destination and relay,
  respectively. Both the relay and destination decode the
  corresponding flows of information at the end of the time slot.
\item During the second time slot, the relay forwards the information
  that it receives in the first time slot to the destination.
  Simultaneously the source sends another flow of information of rate
  $(1-\alpha)R_1^t$ to the destination over the MA channel. The
  destination decodes the two flows of information at the end of the
  time slot.
\end{enumerate}
Note that this flow-oriented method employs only coding for the BC and
MA component channels. No block Markov coding is needed. Compared with
the parallel Gaussian coding in \cite{Host05}, this approach has the
advantage that the destination performs independent decoding at the
end of the two time slots, without the need of storing the received
message in the first time slot for decoding in the second time slot.
This convenient method is applied to the flow optimization design
considered in \cite{Wong07}.

\section{Full-duplex Construction}
Previously the half-duplex relay channel is constructed from the BC
and MA components for the two modes of the relay. Two natural
questions arises from this construction:
\begin{itemize}
\item If the relay can be made to operate in the full-duplex manner,
  what will be the full-duplex relay channel ``physically''
  corresponding to the two components?
\item How does this full-duplex channel compared with the half-duplex
  channel in terms of capacity?
\end{itemize}

To address the above two questions, we need to ``construct'' a
full-duplex relay channel from the BC and MA components. One
physically reasonable construction is to have the full-duplex relay
channel
$(\mathcal{X}_1,\mathcal{X}_2,p_{0}(y,y_1|x_1,x_2),\mathcal{Y},\mathcal{Y}_1)$
satisfying:
\begin{eqnarray}
\sum_{y_1 \in \mathcal{Y}_1} p_{0}(y,y_1|x_1,x_2) &=& p_{t}(y|x_1,x_2)
\label{eqn:fdct} \\
\sum_{y \in \mathcal{Y}} p_{0}(y,y_1|x_1,x_2) &=&
\sum_{y \in \mathcal{Y}} p_{l}(y,y_1|x_1) 
\label{eqn:fdcl}
\end{eqnarray}
for all $(x_1,x_2,y,y_1) \in \mathcal{X}_1 \times \mathcal{X}_2 \times
\mathcal{Y} \times \mathcal{Y}_1$. The requirement in (\ref{eqn:fdct})
clearly states that the constructed full-duplex channel should behave
the same as the MA component if we concentrate on the part from the
source and relay transmitting to the destination. Similarly, the
requirement in (\ref{eqn:fdcl}) forces the constructed full-duplex
channel to act like the BC component if we concentrate on the link
from the source to the relay.

We note that many full-duplex relay channels satisfying
(\ref{eqn:fdct}) and (\ref{eqn:fdcl}) can be constructed.  In
particular, physically degraded (in the sense of \cite{Cover79}) relay
channels can be constructed in this way. Interestingly, all these
physically degraded relay channels so constructed have the same
capacity, which depends only on $p_t(y|x_1,x_2)$ and $p_l(y,y_1|x_1)$.
However it is possible in general to have the constructed full-duplex
channel capacity to be smaller than that of the original half-duplex
channel. This is physically unreasonable and hence implies that we
need an additional constraint in the construction above. It turns out
that the following constraint will force the constructed full-duplex
channel to have a larger capacity:
\begin{equation}
\sum_{y_1 \in \mathcal{Y}_1} p_{l}(y,y_1|x_1) = p_{t}(y|x_1,q)
\label{eqn:fdq}
\end{equation}
where $q$ is the quiet symbol of the relay mentioned before. The
physical interpretation of this constraint is that when the relay
listens, the link from source to the destination is the same that when
the relay sends out the quiet symbol.

\section{Non-degraded Channels and the Gaussian Case}
For half-duplex relay channels that are not physically degraded, like
\cite{Cover79,Host05,Kramer04}, the max-flow min-cut bound in
(\ref{eqn:maxmin}) gives the outer bound in (\ref{eqn:halfbndout}),
which does not coincides with the rate achieved by the decode-forward
coding technique in (\ref{eqn:halfach}). The decode-forward rate in
(\ref{eqn:halfach}) acts as an inner bound on the capacity. It is
conceivable that the compress-forward approach suggested in
\cite{Cover79} can be applied to obtain another inner bound on the
capacity as in \cite{Kramer04,Host05}. The main difficulty of such a
development is that the relay has to determine the listen-transmit
schedule of the current block from the observed symbols in the
previous block, without correctly decoding of the previous message.

When the links among the source, relay, and destination are all
Gaussian channels, we call the resulting relay channel a
\emph{Gaussian} half-duplex relay channel. It is clear that such a
relay channel is characterized by a Gaussian BC channel in the BC mode
and a Gaussian MA channel in the MA mode. More precisely, conditioned
on the event $\{X_3 = l\}$,
\begin{eqnarray}
  Y &=& X_1 + N \nonumber \\
  Y_1 &=& X_1 + N_1
\label{eqn:gaussian1}
\end{eqnarray}
and conditioned on the event $\{X_3 = t\}$,
\begin{eqnarray}
  Y &=& X_1 + X_2 + N \nonumber \\
  Y_1 &=& 0
\label{eqn:gaussian2}
\end{eqnarray}
where $N$ and $N_1$ are independent zero-mean Gaussian noise random
variables with variances $\sigma^2$ and $\sigma_1^2$, respectively.
From (\ref{eqn:gaussian1}) and (\ref{eqn:gaussian2}), the symbol
``$0$'' acts to both the erasure and quiet symbols described in
Section~\ref{sec:model}. In addition, the restriction in
(\ref{eqn:fdq}) is satisfied. Although this Gaussian half-duplex relay
channel is not physically degraded\footnote{If instead of being
  independent of $N_1$ in (\ref{eqn:gaussian1}) and
  (\ref{eqn:gaussian2}), $N = N_1 + N_0$, where $N_0$ is another
  zero-mean Gaussian random noise independent of $N_1$ (needs $\sigma
  > \sigma_1$), then the Gaussian half-duplex relay channel is
  physically degraded.}, the outer bound (\ref{eqn:halfbndout}) and
inner bound (\ref{eqn:halfbndfin}) still apply, with additional power
constraints, $P_1$ and $P_2$, respectively, on the symbols $X_1$ and
$X_2$ for the maximization operation to make sense. Unfortunately, as
suggested in \cite{Kramer04}, the maximizing input distributions for
both bounds are not Gaussian. Finding these maximizing input
distributions is an open problem.

When the relay listen-transmit schedule is deterministic, the
maximizing input distributions for the outer bound is Gaussian and the
maximum bound is given \cite[Prop. 1]{Host05} in the form of
(\ref{eqn:fixed}) with $R_1^l = C\left(\frac{P_1}{\sigma^2}\right)$,
$R_2^l = C\left(\frac{P_1}{\sigma^2} + \frac{P_1}{\sigma_1^2}\right) -
C\left(\frac{P_1}{\sigma^2}\right)$, $R_1^t =
C\left(\frac{(1-\beta)P_1}{\sigma^2}\right)$, and $R_2^t =
C\left(\frac{P_1+P_2+2\sqrt{\beta P_1 P_2}}{\sigma^2}\right) -
C\left(\frac{(1-\beta)P_1}{\sigma^2}\right)$, where the maximum value
is taken over all $0 \leq \alpha, \beta \leq 1$. The function $C(x) =
\frac{1}{2}\log (1+x)$. The block Markov coding (also the parallel
Gaussian channel) inner bound is also given \cite[Prop. 2]{Host05} in
the form of (\ref{eqn:fixed}) with $R_2^l =
C\left(\frac{P_1}{\sigma_1^2}\right) -
C\left(\frac{P_1}{\sigma^2}\right)$ and the other rates remain the
same as above. This form of the bounds suggests once again the use of
the flow-oriented coding method described in
Section~\ref{sec:randvsfixed}. Unfortunately, it turns out that
neither the rate pair $(R_1^l,R_2^l)$ of the outer bound nor that of
the inner bound lies in the achievable rate region of the Gaussian BC
component channel. However the flow-oriented method may still
outperform the decode-forward method in some cases.  We are also
unclear about how this method compared with the compress-forward
method suggested in \cite{Host05}.

\end{document}